\documentclass[journal]{IEEEtran}
\usepackage[table]{xcolor}
\usepackage[utf8]{inputenc}
\usepackage[colorlinks,urlcolor=blue,linkcolor=blue,citecolor=blue]{hyperref}
\usepackage{xspace}
\usepackage{ifthen}
\usepackage[T1]{fontenc}
\usepackage{amsmath,amssymb}
\usepackage{graphicx}
\usepackage{svg}
\usepackage{hyperref}
\usepackage{listings}
\usepackage{xcolor}
\usepackage{caption}
\usepackage{subcaption}
\usepackage{listings}
\usepackage{tabularx}
\usepackage{booktabs, multirow} 
\usepackage{algorithm}
\usepackage{algpseudocode}
\usepackage{numprint}
\usepackage{wrapfig}

\usepackage{tikz}

\usepackage[framemethod=tikz]{mdframed}
\mdfdefinestyle{mpdframe}{
    frametitlebackgroundcolor   =black!15,
    frametitlerule              =true,
    roundcorner                 =1pt,
    middlelinewidth             =1pt,
    skipabove                   =\baselineskip,
    innermargin                 =0.1cm,
    innerleftmargin             =0.1cm,
    innerrightmargin            =0.1cm,
    innertopmargin              =0.1cm,
    innerbottommargin           =0.1cm
}

\usepackage{pifont}

\IEEEoverridecommandlockouts

\colorlet{punct}{red!60!black}
\definecolor{background}{HTML}{EEEEEE}
\definecolor{delim}{RGB}{20,105,176}
\colorlet{numb}{magenta!60!black}

\lstdefinelanguage{json}{
    basicstyle=\footnotesize\ttfamily,
    numbers=left,
    numberstyle=\scriptsize,
    stepnumber=1,
    numbersep=8pt,
    showstringspaces=false,
    breaklines=true,
    frame=lines,
    backgroundcolor=\color{background},
    literate=
     *{:}{{{\color{punct}{:}}}}{1}
      {,}{{{\color{punct}{,}}}}{1}
      {"}{{{\color{punct}{"}}}}{1}
      {\{}{{{\color{delim}{\{}}}}{1}
      {\}}{{{\color{delim}{\}}}}}{1}
      {[}{{{\color{delim}{[}}}}{1}
      {]}{{{\color{delim}{]}}}}{1},
}

\newboolean{showcomments}
\setboolean{showcomments}{true}
\ifthenelse{\boolean{showcomments}}
{
}

\newcommand{\ie}{i.e.\@\xspace}

\newcommand{\eg}{e.g.\@\xspace}
\newcommand{\etal}{et al.\@\xspace}

\newcommand{\geth}{\textsc{Geth}\@\xspace}
\newcommand{\erigon}{\textsc{Erigon}\@\xspace}
\newcommand{\nethermind}{\textsc{Nethermind}\@\xspace}
\newcommand{\besu}{\textsc{Besu}\@\xspace}
\newcommand{\neth}{\textsc{N-ETH}\@\xspace}

\newcommand*\circled[1]{\tikz[baseline=(char.base)]{
    \node[shape=circle, draw, inner sep=1pt, 
        minimum height=12pt] (char) {#1};}}

 \newcommand{\linebreakand}{%
      \end{@IEEEauthorhalign}
      \hfill\mbox{}\par
      \mbox{}\hfill\begin{@IEEEauthorhalign}
    }

\newcommand{\rqone}{What are the behavioral consequences of unstable execution environments for blockchain nodes?}

\newcommand{\rqtwo}{To what extent do different blockchain node versions exhibit different availability rates under unstable execution environments?}

\newcommand{\rqthree}{To what extent does an N-version blockchain node increase availability compared to a single-version node?}

\usetikzlibrary{shapes.geometric}

\newcommand\score{
  \protect\begin{tikzpicture}
      \protect\node[star, 
        star points=5,
        star point ratio=2.25,
        draw,
        inner sep=1pt,
        fill=yellow]{};
  \end{tikzpicture}
}

\newcommand\triup{
\protect\begin{tikzpicture}
  \protect\node[isosceles triangle,
    isosceles triangle apex angle=50,
    draw,
    rotate=90,
    fill=green,
    inner sep=1pt,minimum size=1pt] (T1)at (0,0){};
  \end{tikzpicture}
  }

\title{Highly Available Blockchain Nodes\\ With N-Version Design}

\author{\IEEEauthorblockN{Javier Ron\IEEEauthorrefmark{1}, C\'esar Soto-Valero\IEEEauthorrefmark{1}, Long Zhang\IEEEauthorrefmark{1}, Benoit Baudry\IEEEauthorrefmark{1}, Martin Monperrus\IEEEauthorrefmark{1}}\\
\

\IEEEauthorblockA{\IEEEauthorrefmark{1}\textit{KTH Royal Institute of Technology}}
}

\date{February 2023}
\begin{document}

\maketitle

\thispagestyle{plain}

\pagestyle{plain}

\begin{abstract}
As all software, blockchain nodes are exposed to faults in their underlying execution stack.
Unstable execution environments can disrupt the availability of blockchain nodes' interfaces, resulting in downtime for users.
This paper introduces the concept of N-Version Blockchain nodes. This new type of node relies on simultaneous execution of different implementations of the same blockchain protocol, in the line of Avizienis' N-Version programming vision.
We design and implement an N-Version blockchain node prototype in the context of Ethereum, called \neth.
We show that \neth is able to mitigate the effects of unstable execution environments and significantly enhance availability under environment faults.
To simulate unstable execution environments, we perform fault injection at the system-call level.
Our results show that existing Ethereum node implementations behave asymmetrically under identical instability scenarios.
\neth leverages this asymmetric behavior available in the diverse implementations of Ethereum nodes to provide increased availability, even under our most aggressive fault-injection strategies.
We are the first to validate the relevance of N-Version design in the domain of blockchain infrastructure.
From an industrial perspective, our results are of utmost importance for businesses operating blockchain nodes, including Google, ConsenSys, and many other major blockchain companies. 
\end{abstract}

\section{Introduction}

Blockchain technology is fundamental to offer secure, reliable, and decentralized software services~\cite{masteringETH}.
Blockchains enable the transaction of digital currencies~\cite{BitcoinPaper}, as well as the creation and execution of smart contracts~\cite{Dannen2017}.
Several businesses depend on the correct operation of blockchain networks to provide services to their clients~\cite{Chen2020DeFi}.
Major actors such as banks or cryptocurrency exchanges with high volumes of end-users rely  on trustworthy and uninterrupted access to said networks~\cite{Cocco2017Banks, CoinGecko, Soto2022}.

All those actors exercise their mission-critical business on top of blockchain nodes~\cite{AbouJaoude}.
However, blockchain nodes are never without risk of failure, and blockchain outages have occurred several times, causing downstream service disruptions and loss of revenue~\cite{Outage2020, Outage2022}.
Software failures are often the consequence of operating system, network, or hardware problems, which cause unstable execution environments~\cite{Teng}.
Therefore, there is an overt need for techniques that allow blockchain node operators to mitigate the effects of software failures. 

\textit{N-Version programming} is a proven approach to building fault-tolerant systems~\cite{Avizienis1985}.
N-Version programming consists of creating several implementations called \textit{versions} of a program based on the same specification.
These versions are meant to be executed simultaneously with the same inputs, and the produced outputs are compared afterward as means of fault detection and fault tolerance.

For some blockchains, multiple compatible implementations exist, given that blockchains are protocol-driven by design.
For example, Ethereum's execution and consensus layers have respectively four and five major implementations~\cite{EthereumDocs}.
In general, the usage of many versions of blockchain implementations is regarded as an important addition towards achieving systemic reliability~\cite{ClientDiversity,Yang2021,Finality2023}.

In this paper, our key insight is to take advantage of diverse implementations of blockchain node to enhance dependability properties.
This is realized as the novel concept of “N-Version blockchain node'',
which is an ensemble of diverse blockchain nodes which collectively provide improved services to external clients.
We evidence that N-Version blockchain nodes are a valuable approach for users with high-availability requirements.
To the best of our knowledge, taking advantage of N-Version design for blockchain is novel and has not been proposed before.

To implement our vision of an N-Version blockchain node, we carry out the following steps.
First, we design a novel architecture that provides higher availability to external clients while encapsulating the inner complexity of N-Version software.
This N-Version design includes strategies for request routing, error handling, and response comparison and ranking in the context of blockchain nodes.
Second, we implement an N-Version blockchain node prototype for one of the most sophisticated, feature-rich, complex, and globally adopted blockchains: Ethereum.
We deploy and coordinate several Ethereum client implementations behind a common interface,  in production.
Third, we set up an original experimental framework for measuring blockchain node availability.
This includes characterizing response latency and correctness, designing a comparison oracle in the domain's specific context.
For these experiments, we synthesize realistic fault injection strategies that cause unstable execution environments.
Last, we compute the availability rates of common blockchain nodes as well as the rates of our N-Version prototype.

Our results provide empirical evidence that the behavior diversity of blockchain clients can be harnessed in an N-Version architecture to provide high availability under an unstable execution environment.
Specifically, our prototype  provides higher availability than any of the state-of-the-art nodes taken in isolation.
Under the tested workloads and fault injection strategies, we observe an increase in full availability from $84.7$\% for the best performing single version, to $98.5$\% for the N-Version design.
Our results also show the trade-offs between availability and CPU, disk, and memory usage.

To sum up, our contributions are:
\begin{itemize}
\item The concept of N-Version design for blockchain nodes.
A blueprint of an N-Version blockchain node that leverages the natural diversity of blockchain implementations.
To our knowledge, this is the first-ever realization of the N-Version design vision in the context of blockchain systems. 
\item An implementation of the diverse N-Version blueprint for the Ethereum blockchain using state-of-the-art Ethereum implementations.
\item A sound methodology for studying availability of blockchain nodes in production using realistic fault injection, and the corresponding sound results demonstrating the strong advantages of N-Version for blockchain.
\item A publicly available, open science repository for reproducing our experiments, at \url{https://github.com/ASSERT-KTH/N-ETH/}.
\end{itemize}

\section{Background}
\begin{figure}
\begin{subfigure}{\linewidth}
\centering
\begin{lstlisting}[language=json]
{
  "jsonrpc": "2.0",
  "method": "eth_getBlockByNumber",
  "id": 1,
  "params": ["0xa55e27", false]
}
\end{lstlisting}

\caption{Ethereum JSON-RPC request.}
\label{lst:json-request}
\vspace{2em}
\end{subfigure}

\begin{subfigure}{\linewidth}
\centering
\begin{lstlisting}[language=json]
{
  "jsonrpc": "2.0",
  "id": 1,
  "result": {
    "difficulty": "0xa9ed0e03a6530",
    "gasLimit": "0xbea427",
    "gasUsed": "0xbe64e3",
    "nonce": "0xde84c6458a7c0aa0",
    "number": "0xa55e27",
    "size": "0x6f75",
    "timestamp": "0x5f5ad163",
    "totalDifficulty": "0x3ab7010902da66c075f",
    "transactions": [...],
    "uncles": [],
    ...
  }
}
\end{lstlisting}
\caption{Partial Ethereum JSON-RPC response.}
\label{lst:json-response}
\end{subfigure}
\caption{Ethereum node and external application interaction.}
\end{figure}

\subsection{Blockchain Technology}
\label{sec:ethereum}
Blockchains are distributed ledgers, where data is aggregated and stored in discrete units called blocks~\cite{BitcoinPaper}.
Blockchains are created by \textit{peer-to-peer networks}, where each peer, or \textit{node}, is a host that executes a blockchain client.
A blockchain stored on disk is unlikely to crash until it runs.

\begin{figure}[t]
    \centering
    \includegraphics[width=0.475\textwidth]{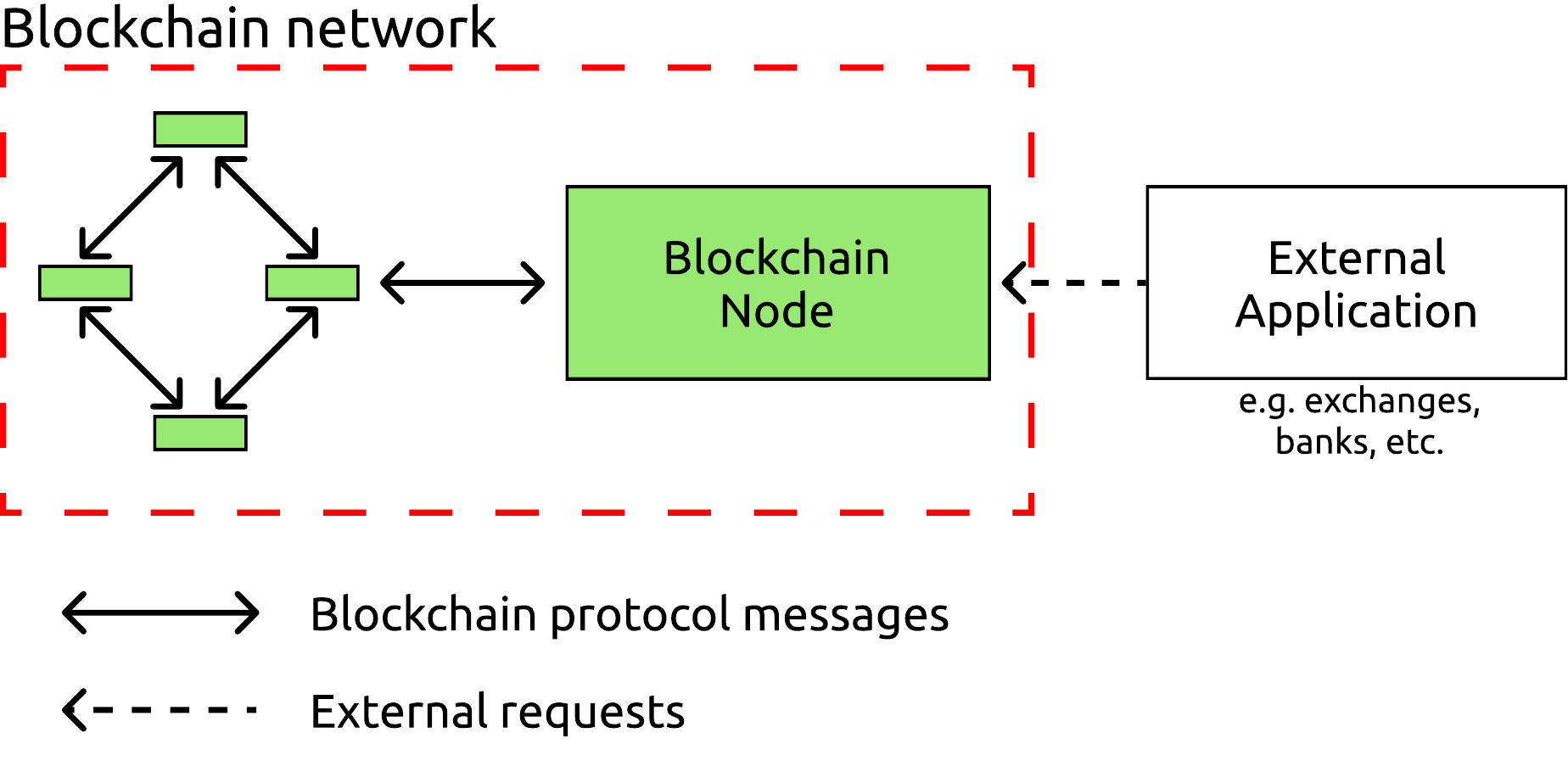}
    \caption{An external application may interact with a blockchain through any node that exposes its interface.}
    \label{fig:single-eth-client}
\end{figure}

Blockchain clients implement a protocol, which describes how blocks are verified, agreed upon, and propagated throughout the peer-to-peer network.
Decentralization is essential to blockchains, therefore every participating node must be able to store a replica of the ledger, and verify all incoming blocks before making them part of their own state.
Ideally, every node also exposes an interface, allowing the blockchain's users to issue read/write queries.

Each blockchain has its own protocol, prime examples of these protocols are Bitcoin's~\cite{BitcoinPaper} and Ethereum's~\cite{ETHYellowPaper}.
All the client implementations for a given blockchain must comply with this blockchain's protocol, and no other restrictions are made on the implementation.
This allows for the creation of several, diverse, client implementations for the most popular blockchains~\cite{BitcoinClients, ClientDiversity}.

In addition to peer-to-peer communication, blockchain nodes provide standardized outward-facing channels to interact with external applications.
For example, Ethereum clients implement the JSON-RPC API specification~\cite{JsonRPC2013}. 
This API allows external applications to connect and query the blockchain using a uniform set of methods.
The available operations include, \eg querying the status of the blockchain at an arbitrary point in the past or issuing new, state-altering transactions.
External applications take advantage of the API to build various services, such as cryptocurrency exchanges, dApps, or games.

\autoref{fig:single-eth-client} depicts an overview of both a blockchain network and an external application. 
Here, two types of interactions are highlighted:
Dotted lines show external applications requests toward a target blockchain node through its outward-facing interface;
and solid lines show the blockchain node which sends and receives data to and from its connected peers.

An example of data exchange between a blockchain node and an external application is shown in Listings \ref{lst:json-request}~and~\ref{lst:json-response}. Listing~\ref{lst:json-request} shows a complete Ethereum JSON-RPC call.
It contains the version of the interface (\texttt{jsonrpc}), the name of the method to be invoked (\texttt{method}), the parameters passed to the method (\texttt{params}), and a request identifier (\texttt{id}).
Listing~\ref{lst:json-response} contains the data returned for the previous request, which is specific to the “\texttt{eth\_getBlockNumber}'' method.
Among other information, it contains the number of the block (\texttt{number}), its size (\texttt{size}), a timestamp (\texttt{timestamp}), and a list of transactions (\texttt{transactions}).

The example in \autoref{fig:single-eth-client} is greatly simplified compared to reality: a production blockchain network is composed of thousands of nodes spread in complex topologies all over the globe~\cite{toposhot}.
Each node serves numerous external applications, meaning that there are many more external applications than nodes.

\subsection{N-Version Design}

An N-Version software application relies on the simultaneous execution of $N$ programs that are different implementations of the same specification~\cite{avizienis1995methodology}.
This type of architecture is used to enhance a desired property of the application, such as reliability~\cite{Yang2021, Tso1986ErrorRI}, performance~\cite{Gholami2020}, or security~\cite{Xu2017, DBLP:conf/dsn/EspinozaWFT22}.
In most cases, this is achieved by comparing or matching the outputs of the $N$ programs given the same input.

N-Version software applications are typically produced through \textit{N-Version programming}.
N-Version programming is defined as the independent development of the same specification by different teams~\cite{chen1978n}.
In this context, each program developed by one team is called a \textit{version}.
Ideally, each version is implemented independently using competing designs, programming languages, and software stacks~\cite{Avizienis1985}.
A high degree of diversity is key, as it lowers the probability of shared faults between versions~\cite{Zavala2008, Voas:1997:ReducingUncertaintyAboutCMF, Townend2001}.

Interestingly, there exist software specifications for which multiple isolated implementations surface \textit{naturally}~\cite{Baudry2015}. 
These implementations emerge with no coordinated effort to produce them.
Instead, they emerge spontaneously due to market competition, the need for optimization, or opinionated design approaches~\cite{sotoDeps}.
Web browsers are an example of this pattern: they are developed independently of each other by competing actors, and yet they conform to the same standards~\cite{DBLP:conf/ndss/XueDK12}.

When independent versions that comply to the same protocol emerge naturally, it is possible to harness them to build N-Version software applications~\cite{Baudry2014}. This approach can be called \textit{natural N-Version design}. 

\textbf{Definition:} Natural N-Version design is the creation of N-Version software applications by harnessing, deploying, and simultaneously executing already-available implementations of a software specification.

\section{Highly Available Blockchain Node}
\subsection{Blockchain Node Availability}
\label{sec:availability}
The availability of a blockchain can be defined as the probability that it is functioning correctly at an arbitrary point in time~\cite{tradeoffs}.
For the purpose of this paper, we include in this definition the possibility of single blockchain nodes to participate in the network in a degraded state.
Therefore, we characterize blockchain nodes' availability as a categorical variable, which can hold three values:
\textit{Available:} The API's responses are timely, compliant, and fresh;
\textit{Degraded:} The API's responses are not compliant or not fresh; 
and \textit{Unavailable:} The API's responses are not timely or requests are denied.

The properties of the responses are defined as follows.
\textit{Timeliness} refers to obtaining a response to a request within a time span $T$. For example, if $T$ is set to 100ms, a response time $t_r$ of 150ms is not timely.
\textit{Compliance} refers to a response's conformity to the API schema. For example, truncated, non-parsable, or otherwise defective responses are not compliant. \textit{Freshness} refers to how up-to-date the contents of the response are with respect to the global blockchain state.
It is measured by computing the distance between the latest block available from the API and an external oracle, namely, another blockchain node.
If the freshness value $f_r$ of a response is higher than a maximum freshness value $F$, we consider the response not fresh.


We define the availability status $S$  of a blockchain node $n$ with response $r$ as follows:
\vspace{1pt}
\begin{equation}
S=\begin{cases}
    \textsc{available};& t_r \leq T \land c_r\equiv true \land f_r \leq F \\
    \textsc{degraded};& t_r \leq T \land (c_r\equiv false \lor f_r > F) \\
    \textsc{unavailable};&  t_r > T
\end{cases}
\label{eq:availability}
\end{equation}
\vspace{1pt}

Where $t_r$ is the response time, $c_r$ the compliance of the response, and $f_r$ the freshness of the response.
We measure $t_r$ as time in milliseconds, $c_r$ as a boolean value, and $f_r$ as the block distance between $n$ and an external oracle.
The upper bounds for response time and block distance are $T$ and $F$, respectively, and are set by the node's operator according to their own requirements.

Given that an individual node response defines the status of the node at one point in time, we consider the node to hold that status until it sends a new response.

\subsection{Architecture of an N-Version Blockchain Node}
\label{sec:system-design}
\begin{figure}
    \centering
    \includegraphics[width=0.50\textwidth]{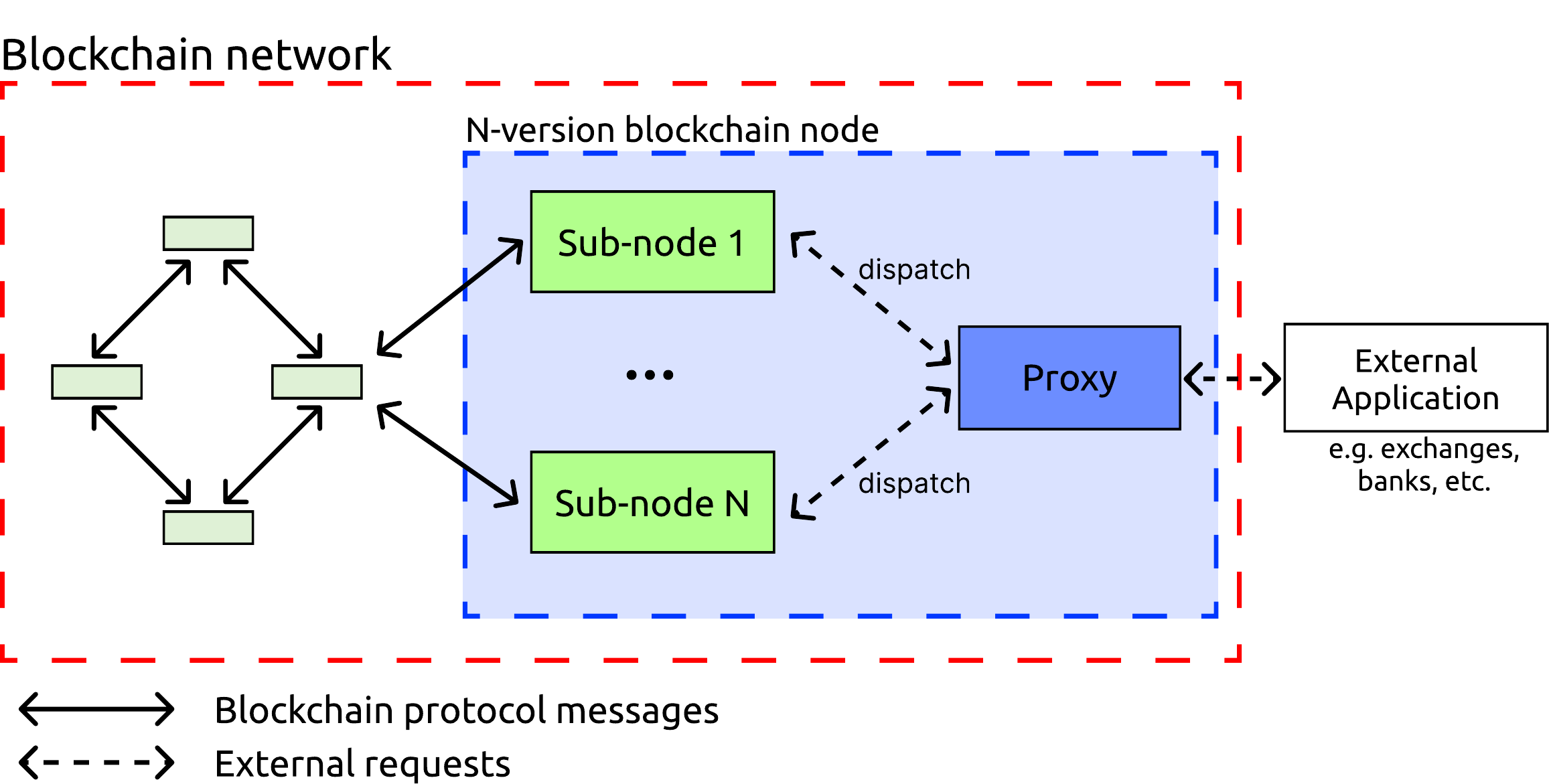}
    \caption{Concept of an N-Version blockchain node, taking advantage of the presence of multiple versions for the same blockchain network.}
    \label{fig:N-Version-eth-client}
\end{figure}

For some blockchains, there are several compatible client implementations~\cite{chainsafe}.
For example, Ethereum's execution layer has four major implementations.
Our goal is to build on this favorable property of blockchains, and apply N-Version design in this context, as shown in \autoref{fig:N-Version-eth-client}.
We call the resulting construct an \textit{N-Version blockchain node}.

\textbf{Definition:} 
An N-Version blockchain node is an ensemble of $N$ sub-nodes and a proxy, where a sub-node is a normal node executing a unique client implementation;
and the proxy encapsulates the sub-nodes under a single interface.

Our study focuses on assessing the impact of  N-Version nodes on availability.
The goal is to demonstrate that an N-Version blockchain node provides higher availability than regular nodes in isolation, specially under unstable execution conditions.


\subsubsection{Overview}

\autoref{fig:N-Version-eth-client} shows a blueprint of the proposed N-Version blockchain node.
First, it shows the components presented in \autoref{sec:ethereum}: Blockchain network, external application, and peer-to-peer communication.
The key novelty in our architecture is the proxy component.
This component exposes an interface which encapsulates $N$ blockchain sub-nodes, where each sub-node executes a different implementation of the blockchain protocol.
Each request directed to the N-Version node must be done through this proxy.

The proxy is responsible for the orchestration of the N-Version node, by routing requests to the sub-nodes depending on a dynamic priority policy and fail-retry mechanisms as explained in \autoref{sec:dispatching}.
The proxy is also in charge of deciding which response to return to the caller in cases where several responses for a single request are produced (\autoref{sec:comparison}). 

\subsubsection{Dispatching Policy}
\label{sec:dispatching}
The presence of sub-nodes  is an opportunity to have an adaptive dispatching policy based on their observed behavior.
Such a policy allows the system to dynamically adjust to the effects of unstable execution environments, by prioritizing the most available sub-node.
This is alike dynamic load balancing, enabling us to achieve system-wide optimization using the global  state of the system~\cite{LoadBalancing}.

The policy works as follows.
We keep an availability score for each sub-node.
The score is the percentage of successful responses for all requests sent to that sub-node.
The score is updated every time a response is received from a sub-node.
With this score, we keep a ranking of the sub-nodes.
Every time the scores are updated, the ranking is sorted in descending order of availability score.
When the proxy receives a request, it is forwarded to the top sub-node in the ranking.
If an \textsc{available} response is returned, said response is sent immediately to the requester.
Otherwise, the proxy saves the response, and retries the request on the next sub-node of the ranking.
This process is repeated until either one of the sub-nodes responds with an \textsc{available} response, or all of the sub-nodes have provided one response.

\subsubsection{Comparison Oracle}
\label{sec:comparison}
When no sub-node is fully available, there is a need to select the best degraded response to be returned to the external application. 
For this, the system compares the sub-noed responses and sends back the best one according to the following rules.
\textit{Rule~1:} A compliant response is better than a non-compliant response; and 
\textit{Rule~2:} The most fresh response from all compliant responses is better.
Compliance is used as a primary filter, because non-compliant responses can cause undefined downstream behavior.
For instance, a response with an incomplete JSON object, could trigger unhandled errors on the external caller.

\subsection{Implementation}

We fully implement a prototype based on the blueprint architecture described in \autoref{sec:system-design} in the context of the Ethereum blockchain.
We call this prototype \neth.
We use the readily available implementations of Ethereum execution-layer clients \geth v1.12.2, \besu v23.7.0, \erigon v2.48.1, and \nethermind v1.20.1 as the sub-nodes of the system.
Each of the chosen implementations has an active community and is open source.
In the rest of this paper, we refer to them as \textit{Ethereum node versions}.
The proxy component is written in Go, using networking components from the standard library.

\begin{figure*}[h]
\centering
\begin{subfigure}[c]{0.44\textwidth}
    \includegraphics[width=1\textwidth]{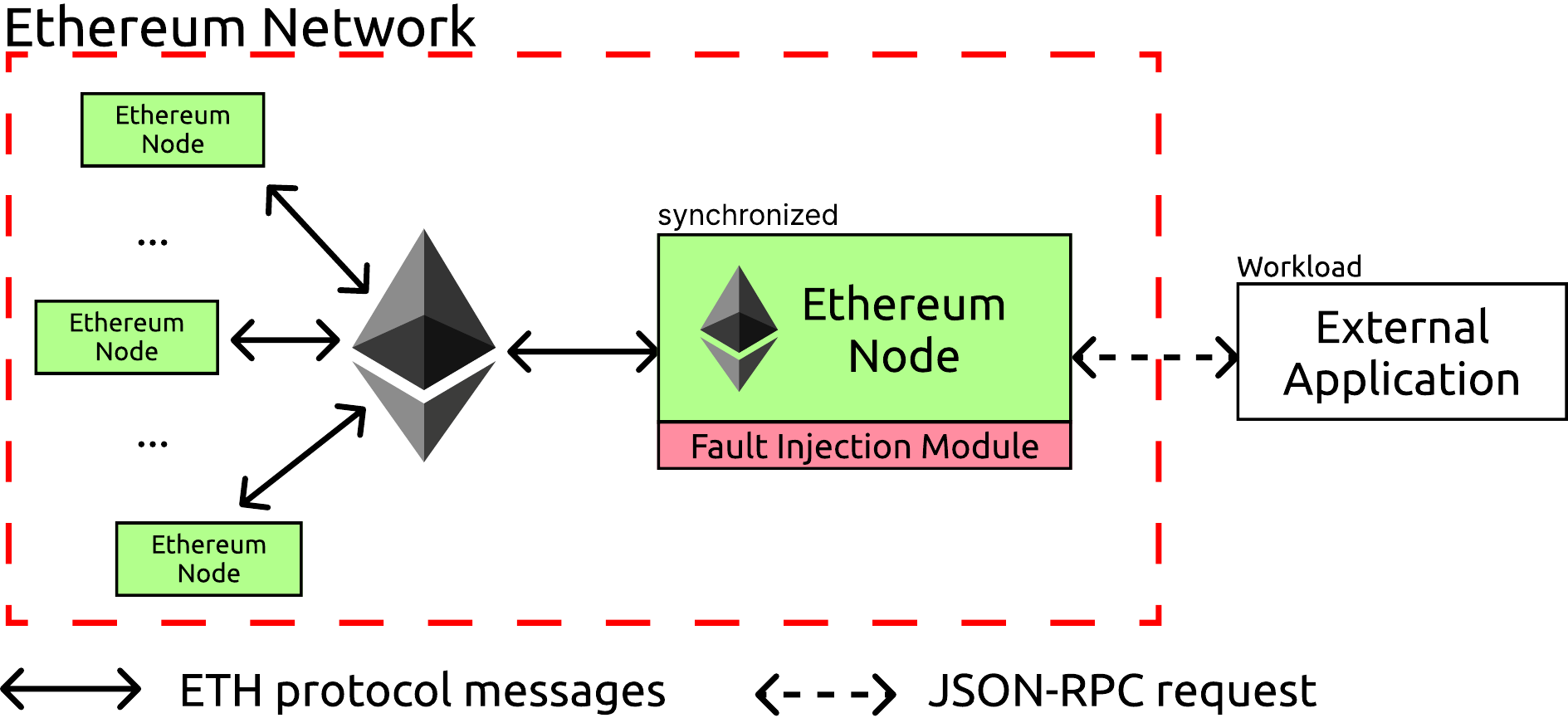}
    \subcaption{Single-version deployment}
    \label{fig:single-version-eth}
\end{subfigure}
\hfill
\begin{subfigure}[c]{0.54\textwidth}
    \includegraphics[width=1\textwidth]{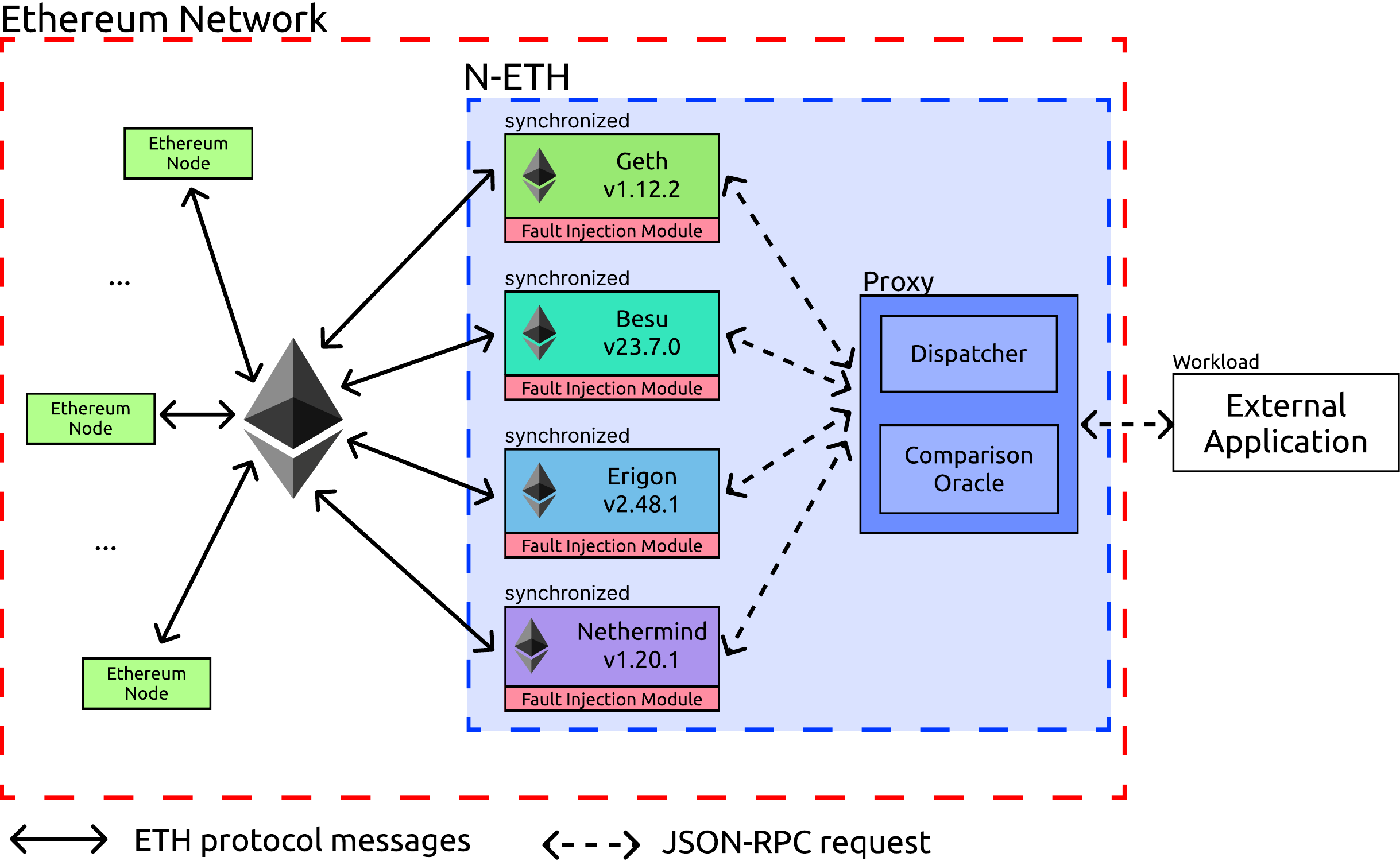}
    \subcaption{N-Version deployment (N=4)}
    \label{fig:N-Version-eth}
\end{subfigure}
\caption{Overview of the experimental Ethereum deployments. Each node or sub-node takes 8–16 hours to synchronize with Ethereum's Mainnet. In both figures, the Ethereum logo represents Ethereum's Mainnet, which comprises tens of thousands of nodes.}
\end{figure*}

\section{Experimental Protocol}

\subsection{Research Questions}

To systematically evaluate our architecture and prototype, we propose the following research questions:
\\

\noindent\textbf{RQ1}~\textit{\rqone}

Blockchain nodes may behave incorrectly due to unstable execution environments.
We aim to identify the effects caused by this instability.
To this end, we deploy four Ethereum nodes, each with system-call error injection in place.
By varying the fault injection strategies, we observe and record a wide range of irregular behavior that may have an impact on availability.
\\

\noindent\textbf{RQ2}~\textit{\rqtwo}

To establish a baseline for availability, we perform a quantitative analysis of the identified effects.
To obtain this baseline, we deploy single Ethereum nodes, each with a corresponding system-call error injection module.
The baseline consists of precise measurements of the availability state of the nodes under increasingly aggressive fault injection strategies.
The availability state is recorded as defined in \autoref{eq:availability}.
\\

\noindent\textbf{RQ3}~\textit{\rqthree}

We argue that an N-Version blockchain node enhances availability properties of the node under unstable execution environments.
To measure this improved availability, we deploy our N-Version Ethereum node prototype with attached system-call error injection modules.
We measure its availability score with varying N and compare it against the baseline derived from single-version nodes.
\\

\subsection{Deployments}
To answer the research questions, we deploy single-version Ethereum nodes and \neth nodes, on which we exert fault injection strategies and workloads.
When selecting the versions that constitute the N-Version deployments, we can only select among existing Ethereum node versions.
According to the Ethereum community, there are 4 implementations that make up for virtually all participating nodes in the main network~\cite{ClientDiversity}.
We use these 4 node implementations as part of our experimental setup: \geth, \erigon, \besu, and \nethermind.

\textit{Single-version deployment:} \autoref{fig:single-version-eth} shows the scope of the single-version deployment.
\label{sec:single-version-deployment}
To realize it, we first deploy a particular Ethereum node version and configure it to synchronize with Ethereum's Mainnet.
Once synchronized, we attach a fault injection module to the Ethereum node process, as explained below, in \autoref{sec:fault-injection-strategies}.
The data collected from this deployment provides insights into RQ1 and RQ2, \ie to identify and measure the effects of unstable execution environments in the nodes' availability.

\textit{N-Version deployment:} 
\label{sec:N-Version-deployment}
To realize an N-Version deployment, we go through the following steps:
First, we create $N$ instances of Ethereum nodes, each coming from a different version.
We configure them to synchronize with Ethereum’s Mainnet.
Second, we deploy an instance of the proxy and connect it with those synchronized nodes as sub-nodes.
Third, we attach a fault injection module to each of the sub-nodes.
We perform these three steps with N equal to 2, 3, and 4.
\autoref{fig:N-Version-eth} shows the scope of \neth~with $N=4$: the proxy is connected to 4 subnodes, an instance of Geth, an instance of Besu, an instance of Erigon, and an instance of Nethermind.
For each value of N, we consider all possible sub-node combinations.
In total this adds up to 29 deployments.
The data collected from these deployments provides insights into RQ3, \ie measuring the improved availability under unstable execution environments.

In both single-version and N-Version deployments, we record the availability state for each request, which can take three different values: \textsc{available}, \textsc{degraded}, or \textsc{unavailable}, as described in \autoref{sec:availability}.
Additionally, we measure the resource consumption of each deployment, to determine the tradeoff between N and any change of measured availability.

\subsection{Workloads}
\label{sec:workload}
Workloads are exerted into the deployments through a custom component, which acts as the 'external application' component depicted in \autoref{fig:single-version-eth} and \autoref{fig:N-Version-eth}.
The workloads consist of an arbitrary number and types of JSON-RPC method invocations targeting the deployment.
To quantify the availability of deployment, the workload component records the received responses' conformity, freshness, and latency.
We devise two workloads:

\textit{Workload A} consists of \numprint{360000} JSON-RPC method invocations, where each invocation's method name and parameters are sampled from a pool. This pool contains 21 methods and corresponding parameters, which query both current and past states of the blockchain. The aim of this workload is to discover the widest possible range of availability-related issues induced by our fault injection strategies. 

\textit{Workload B} consists of \numprint{360000} invocations of a single JSON-RPC method, which queries for the latest block available on the target deployment.
The aim of this workload is to collect freshness information.
It is important to note that while the requests are identical, the responses are expected to change regularly over time as more blocks are added to the chain.

Both workloads are configured to perform each request $5$ milliseconds apart.
This means that in total, the workloads will perform requests steadily for 30 minutes.
Over this time span, the Ethereum blockchain adds 150 blocks to the chain.
The selected workload duration and ensuing block distance allow us to observe the effects of unstable execution environments on our deployments.

\subsection{Unstable Environment Simulation}
\label{sec:fault-injection-strategies}

Blockchain nodes are executed on top of an operating system (OS).
Consequently, blockchain nodes are susceptible to OS or hardware instability.
In environments such as the cloud, single hardware or network faults can propagate to multiple virtual machines~\cite{correlated-failures} affecting multiple blockchain client instances simultaneously.
Such instability typically manifests downstream as system-call invocation errors~\cite{Forrest1996}.
For example, a \texttt{read} system-call may repeatedly fail with error code \texttt{-EAGAIN} due to a disk malfunction. 
Previous research shows that high-frequency system-call errors may cause unexpected behavior~\cite{Zhang2021phoebe}.
In the context of blockchains, it can result in disruption of block transmission, and chain synchronization~\cite{Zhang2021chaosETH}.
Additionally, permanent side effects and crashes can also be the result of system-call errors. 
Therefore, we consider system-call errors as the fault model under which we analyze the degradation of blockchain client APIs.
Fault models that can make blockchain nodes unavailable with 100\% certainty, such as power outages, are out of the scope of this study.

\begin{figure}
\centering
    \includegraphics[width=0.475\textwidth]{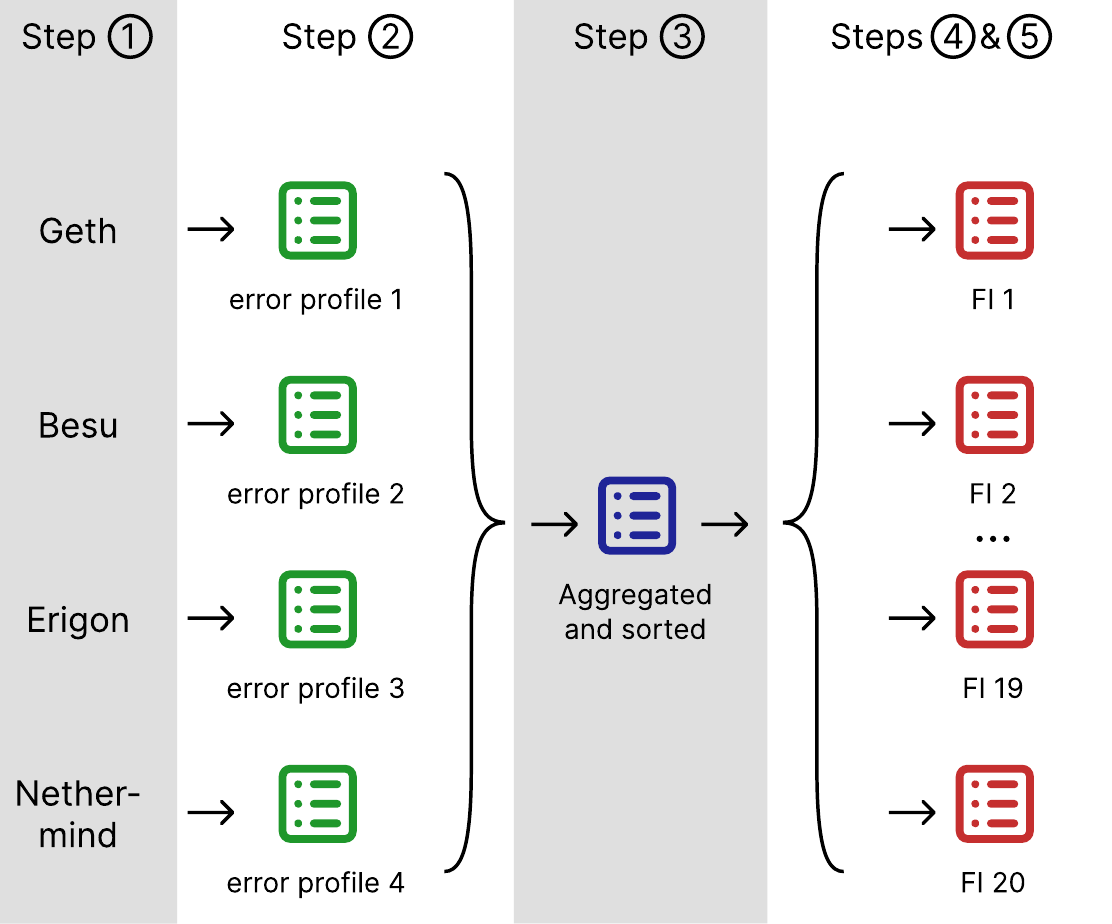}
    \caption{Fault injection strategy synthesis.}
    \label{fig:fi-strategies}
\end{figure}

Since fault injection is regarded as an effective way to test N-Version systems~\cite{DBLP:conf/words/TownendX02}, we devise several realistic fault injection strategies~(FIs) to apply into blockchain nodes.
\autoref{fig:fi-strategies} shows the process used to craft realistic fault injection strategies.
It consists of the following steps:
\circled{1} For all the Ethereum node versions used in our deployments, we perform a monitoring procedure, where we record all system-calls and system-call return codes.
The return codes reveal system-call invocations which fail.
Unsuccessful system-calls are frequent even during correct execution of processes,
and may be caused \eg by temporarily unavailable resources or lost connections.
\circled{2} We produce a system-call error profile in the form of a set $S$ of tuples with the form $\langle syscall, err, f \rangle$, where $syscall$ is the name of a system-call, $err$ is a system-call return code, and $f$ is the frequency with which $syscall$ returns with code $err$.
\circled{3} We aggregate the sets of each analyzed version into a single set, where no pair of $syscall$ and $err$ is repeated and $f$ is the minimum value between any pair that was repeated.
The aggregated set is then sorted in descending order based on $f$
\circled{4} We create $n$ subsets from the aggregated set following a top-n pattern, \ie, subset 1 contains the top-1 tuple from the aggregated set, subset 2 contains the top-2 tuples from the aggregated set, and so on. 
\circled{5} In all tuples of the resulting sets, $f$ is amplified with an arbitrary factor of $5\%$.
We consider this factor to result in balanced scenarios where sporadic errors are likely to be observed, while the relative frequency of system-call errors is kept.

After applying these steps, we obtain 20 fault injection strategies with the following attributes:
(1) They are realistic, this means that they only include system-call errors known to occur spontaneously in at least one of the analyzed blockchain nodes;
(2) They generate faults uniformly and independent of the deployments' nodes or sub-nodes, and therefore allows comparing resilience between deployments;
(3) There is a clear increase of aggressiveness from FI~1 to FI~20.
FI~1 contains a single tuple, which is the one with the highest $f$.
We consider the most frequent system-call errors to be handled with high certainty, therefore we regard FI~1 as the least aggressive strategy.
On the other hand, FI~20 contains all 20 observed error tuples. 
This means that FI~20 is the most aggressive strategy, which triggers the highest number of errors, including the most uncommon ones.

To monitor the Ethereum clients' system-calls and to perform fault injection, we rely on the tool ChaosETH~\cite{Zhang2021chaosETH}. 

\subsection{Running Experiments at Scale}

For our experiments to be the closest to a real-world setting, we require them to be done over the main network of Ethereum, called “Mainnet''. Consequently, all nodes  must be fully synchronized with Ethereum's Mainnet.
However, synchronizing an Ethereum node on Mainnet can be challenging~\cite{runnode}.

First, it requires a significant amount of resources:
The selected node versions require from 0.8~TBs to 1.2~TBs of space on fast storage devices, as well as between 8~GBs and 16~GBs of RAM. 
Second, it requires the parallel execution of another kind of blockchain node, known as a \textit{consensus layer} node.
Third, synchronizing an Ethereum node from scratch takes from 10+ hours to several days.
The time largely depends on the hardware where the node is executed, and the available bandwidth.

We devise experiments as follows:
Regarding single-version deployments, we perform one experiment per Ethereum node version (4), per fault injection strategy (20), per workload (2).
Regarding the N-Version deployment, we perform one experiment per fault injection strategy (20), per value of $N$, and corresponding combinations.
Setting all experiments ultimately amounts to executing and synchronizing 660 Ethereum node instances.

To handle this scale, we implement a cloud pipeline that allows us to replicate nodes efficiently.
The pipeline uses a cloud computing setup which provides access to $n > 1$ SSD devices.
The first SSD is reserved for a node instance which is always kept up to date and where no fault injection is performed.
We call this instance the \textit{source} node, and it is synchronized from scratch.

\begin{algorithm}
\caption{Experiment pipeline}\label{alg:cap}
\begin{algorithmic}[1]
\State{$w \gets \{A | B\}$} \Comment{Workload}
\State{$d \gets \{single | neth\}$} \Comment{Deployment}
\State{$S \gets \{S_1 ... S_{20}\}$} \Comment{Fault injection strategies}
\Statex

\Function{main}{}
\State{\textsc{sync\_source($d$)}}
\State{\textbf{wait} $source\_is\_synced$}
\For {$s \leftarrow 0, len(S)$}
\State{\textsc{run\_experiment($d, w, s$)}}
\EndFor
\State{\textbf{wait} $all\_experiments\_finished$}
\State{\textsc{exit ()}}
\EndFunction
\Statex

\Function{sync\_source}{$d$}
\While{$true$}
\State{\textsc{start\_sync ($d$)}}
\State{\textbf{wait} $copies\_in\_progress$}
\State{\textsc{stop\_sync ($d$)}}
\State{\textbf{wait not} $copies\_in\_progress$}
\EndWhile
\EndFunction
\Statex

\Function{run\_experiment}{$d, w, s$}
\State{\textbf{wait} $available\_ssd$}
\State{\textsc{copy\_state()}}
\State{\textsc{start\_deployment($d, s$)}}
\State{\textsc{run\_workload($w$)}}
\EndFunction
\end{algorithmic}
\label{lst:pipeline}
\end{algorithm}

The pipeline then continues by executing three asynchronous procedures, as detailed in Algorithm~\autoref{lst:pipeline}.
\textsc{main} starts the initial source node synchronization by calling \textsc{sync\_source}, waits for the source node to be up-to-date, and finally calls one instance of \textsc{run\_experiment} for each fault injection strategy.
We must keep a source node to later create copies of its state for use in the subsequent experiments.
This means that synchronization from scratch only happens once. 
\textsc{sync\_source} runs an infinite loop, which starts the source node's synchronization, and pauses it while state copying is in progress.
This procedure also restarts source node's synchronization if no copying actions are in progress.
Stopping the source node's synchronization is necessary, since the state of the node changes constantly with each added block.
Keeping the source node constantly synchronized results in the experiments being carried out with the latest production state.   
Finally, \textsc{run\_experiment} copies the state from the source node's SSD to a newly provisioned SSD, and then starts a deployment which includes the blockchain node, fault injection module, and external application.
Then, it starts the experiment's workload.
Performing each experiment on a deployment with a clean copy of the deployment's state guarantees that no lingering effects from fault injection are carried from experiment to experiment.
This novel pipeline is designed for paralellization, and enables us to carry out the experiments in an acceptable timeframe.

The experiments are executed in Microsoft Azure virtual machines of type L64s v3, each of which provides 64 vCPUs, 512 GBs of RAM, and access to 8x 1.8 TBs NVMe SSDs.
This type of environment fits our use case perfectly, since each instance allows us to simultaneously execute up to 8 Ethereum nodes.
The estimated total cost of performing the experiments on the Azure platform is \numprint{10000} USD.

\section{Experimental Results}
\subsection{\rqone}
\autoref{tab:rpc-errors} shows all the error types received by an external application while the target blockchain nodes are under fault injection.
Each row represents the sum of observed errors for all fault-injection strategies, per each single-version deployment, under workload A.
We observe that unstable execution environments have different visible effects on blockchain nodes.
Specifically, 17 different types of errors surface on the external application, and are related to either network issues, timeouts, or data corruption.
Overall, the most frequent type of error is “connect: connection refused”. We determine under log analysis that this error arises when a request is directed towards a crashed deployment.
On the other hand, data corruption errors such as “malformed HTTP response”, or “unexpected end of JSON”, happen very rarely and represent only a small fraction of the total error count.

Regarding the distribution of errors, not all error types have the same frequency in every node version, \ie there are errors which are common in some versions, but rare in others.
For example, the error “Post: Client.Timeout while waiting headers'', occurs with all versions, however its absolute frequency in each deployment varies by orders of magnitude.
Furthermore, the error “invalid character in response'' is very frequent in the \besu deployment, but is never triggered in the rest of the deployments.
Finally, we observe that within the same deployment, the frequency of errors is not distributed uniformly, and the absolute frequencies of errors range from zero to hundreds of thousands.

\begin{table}[!htp]
\centering
\scriptsize
\rowcolors{2}{gray!12}{white}
\setlength{\tabcolsep}{3.5pt}
\begin{tabular}{lrrrr}
\toprule
Error &  \textsc{Ge.} &  \textsc{Be.} &  \textsc{Er.} &  \textsc{Ne}. \\
\midrule
Post: EOF                                   & \numprint{56}     & \numprint{129713} & \numprint{31}     & \numprint{48} \\
Post: Client.Timeout while awaiting headers & \numprint{4820}   & \numprint{357864} & \numprint{3400}   & \numprint{45932} \\
connect: connection refused                 & \numprint{170409} & \numprint{70623}  & \numprint{201038} & \numprint{534307} \\
dial tcp: connect: connection reset by peer & -                 & -                 & -                 & \numprint{1} \\
http: server closed idle connection         & -                 & -                 & -                 & \numprint{2} \\
malformed HTTP response                     & -                 & \numprint{9}      & -                 & - \\
read: connection reset by peer              & \numprint{854}    & \numprint{598}    & \numprint{6153}   & \numprint{9263}\\
read: Client.Timeout while awaiting headers & -                 & -                 & \numprint{1}      & - \\
Client.Timeout while reading body           & \numprint{3}      & -                 & -                 & \numprint{1} \\
gzip: invalid checksum                      & -                 & \numprint{865}    & -                 & - \\
invalid byte in chunk length                & -                 & \numprint{2599}   & -                 & - \\
invalid character in response               & -                 & \numprint{293867} & -                 & - \\
unexpected EOF                              & \numprint{4}      & \numprint{2}      & \numprint{3}      & \numprint{3373}\\
unexpected end of JSON input                & -                 & \numprint{1}      & -                 & - \\
\bottomrule
\end{tabular}
\caption{Count of error types occurrences when running Workload A under all fault injection strategies. (\textsc{Ge.}) \geth, (\textsc{Be.}) \besu, (\textsc{Er.}) \erigon, (\textsc{Ne.}) \nethermind.}
\label{tab:rpc-errors}
\end{table}

There are  5 error types in total, which are triggered only in one node version.
These observations suggest that the type and prevalence of errors are non-coincidental, meaning that the same injected fault triggers vastly different errors, depending on the node version.
This is fully in line with the core N-Version design assumption: diverse implementations exhibit diverse errors.

\begin{table*}
\centering
\scriptsize
\rowcolors{6}{gray!10}{white}
\begin{tabular}{lrrr|rrr|rrr|rrr}
\toprule
{} & \multicolumn{3}{c}{\geth} & \multicolumn{3}{c}{\besu} & \multicolumn{3}{c}{\erigon} & \multicolumn{3}{c}{\nethermind} \\
\cmidrule(lr){2-4} \cmidrule(lr){5-7} \cmidrule(lr){8-10} \cmidrule(lr){11-13}
RPC method name & \multicolumn{1}{c}{FI 19} & \multicolumn{1}{c}{FI 18} & \multicolumn{1}{c}{FI 17} &  
 \multicolumn{1}{c}{FI 19} & \multicolumn{1}{c}{FI 20} & \multicolumn{1}{c}{FI 17} & 
 \multicolumn{1}{c}{FI 20} & \multicolumn{1}{c}{FI 18} & \multicolumn{1}{c}{FI 19} &
 \multicolumn{1}{c}{FI 20} & \multicolumn{1}{c}{FI 17} & \multicolumn{1}{c}{FI 18} \\
\midrule
eth\_blockNumber                         & 0.138 & 0.184 & 0.068 & 0.377 & 0.022 & 0.835 & 0.159 & 0.144 & 0.139 & 0.366 & 0.377 & 0.424 \\
eth\_estimateGas                         & 0.140 & 0.182 & 0.069 & 0.376 & 0.023 & 0.834 & 0.163 & 0.147 & 0.131 & 0.378 & 0.377 & 0.423 \\
eth\_feeHistory                          & 0.138 & 0.183 & 0.065 & 0.384 & 0.023 & 0.839 & 0.161 & 0.141 & 0.138 & 0.369 & 0.372 & 0.423 \\
eth\_gasPrice                            & 0.137 & 0.182 & 0.069 & 0.382 & 0.024 & 0.840 & 0.160 & 0.143 & 0.138 & 0.369 & 0.373 & 0.427 \\
eth\_getBalance                          & 0.138 & 0.182 & 0.066 & 0.382 & 0.022 & 0.837 & 0.157 & 0.147 & 0.136 & 0.368 & 0.376 & 0.421 \\
eth\_getBlockByHash                      & 0.136 & 0.188 & 0.066 & 0.392 & 0.032 & 0.843 & 0.158 & 0.144 & 0.139 & 0.375 & 0.369 & 0.421 \\
eth\_getBlockByNumber                    & 0.131 & 0.186 & 0.067 & 0.390 & 0.028 & 0.840 & 0.161 & 0.145 & 0.142 & 0.373 & 0.377 & 0.425 \\
eth\_getBlockTransactionCountByHash      & 0.141 & 0.188 & 0.066 & 0.382 & 0.023 & 0.839 & 0.159 & 0.142 & 0.136 & 0.365 & 0.376 & 0.425 \\
eth\_getBlockTransactionCountByNumber    & 0.139 & 0.185 & 0.067 & 0.384 & 0.022 & 0.841 & 0.161 & 0.146 & 0.134 & 0.375 & 0.368 & 0.423 \\
eth\_getCode                             & 0.140 & 0.183 & 0.068 & 0.389 & 0.028 & 0.843 & 0.157 & 0.143 & 0.140 & 0.367 & 0.375 & 0.427 \\
eth\_getLogs                             & 0.144 & 0.187 & 0.061 & 0.400 & 0.039 & 0.850 & 0.162 & 0.147 & 0.140 & 0.368 & 0.381 & 0.420 \\
eth\_getStorageAt                        & 0.139 & 0.184 & 0.065 & 0.379 & 0.022 & 0.838 & 0.156 & 0.145 & 0.138 & 0.368 & 0.374 & 0.427 \\
eth\_getTransactionByBlockHashAndIndex   & 0.142 & 0.188 & 0.063 & 0.387 & 0.022 & 0.838 & 0.157 & 0.150 & 0.137 & 0.367 & 0.371 & 0.420 \\
eth\_getTransactionByBlockNumberAndIndex & 0.135 & 0.184 & 0.063 & 0.385 & 0.023 & 0.844 & 0.163 & 0.144 & 0.136 & 0.371 & 0.370 & 0.421 \\
eth\_getTransactionByHash                & 0.140 & 0.188 & 0.067 & 0.387 & 0.026 & 0.834 & 0.184 & 0.166 & 0.154 & 0.372 & 0.374 & 0.418 \\
eth\_getTransactionCount                 & 0.141 & 0.181 & 0.067 & 0.381 & 0.024 & 0.838 & 0.154 & 0.145 & 0.136 & 0.375 & 0.370 & 0.419 \\
eth\_getTransactionReceipt               & 0.139 & 0.179 & 0.069 & 0.386 & 0.024 & 0.837 & 0.176 & 0.161 & 0.155 & 0.372 & 0.379 & 0.419 \\
eth\_getUncleByBlockHashAndIndex         & 0.143 & 0.184 & 0.068 & 0.391 & 0.023 & 0.839 & 0.157 & 0.144 & 0.140 & 0.374 & 0.372 & 0.419 \\
eth\_getUncleByBlockNumberAndIndex       & 0.138 & 0.182 & 0.065 & 0.377 & 0.024 & 0.835 & 0.158 & 0.140 & 0.138 & 0.362 & 0.379 & 0.427 \\
eth\_getUncleCountByBlockHash            & 0.139 & 0.181 & 0.065 & 0.382 & 0.023 & 0.839 & 0.156 & 0.146 & 0.137 & 0.366 & 0.377 & 0.421 \\
eth\_getUncleCountByBlockNumber          & 0.135 & 0.185 & 0.067 & 0.385 & 0.022 & 0.840 & 0.159 & 0.144 & 0.140 & 0.372 & 0.373 & 0.425 \\
\midrule
SD                                       & 0.003 & 0.003 & 0.002 & 0.006 & 0.004 & 0.004 & 0.007 & 0.006 & 0.006 & 0.004 & 0.003 & 0.003 \\
\bottomrule
\end{tabular}
    \caption{Error rate per method, client, and fault injection strategy. Workload A.}
    \label{table:rpc-availability-single}
\end{table*}

\autoref{table:rpc-availability-single} shows the proportion of RPC calls which triggered an error in the external application.
The first column contains the JSON-RPC method names, and from then on, each column presents the results for each deployment under three fault injection strategies and workload A.
For instance, the cell at the intersection of “eth\_blockNumber'', \geth, and FI~18, indicates that $13.8$\% of the requests for this RPC-deployment-FI combination cause an error observed in the external application.

The bottom row of \autoref{table:rpc-availability-single} shows the standard deviation (SD) for each column.
Analyzing the SD values, we find that the effects of the fault injection strategies on the different methods of the APIs are highly uniform.
The column with highest SD corresponds to \erigon under FI~20, with a value of $0.007$.
This means that the tested fault injection strategies do not have significantly varying effects depending on the measured API method, which is a good indicator of external validity.
For simplicity, \autoref{table:rpc-availability-single} presents only the results of three fault injection strategies for each node version, the ones with the largest SD.
The complete information for all RPC-deployment-FI combinations is available at \url{http://github.com/ASSERT-KTH/N-ETH}

\begin{table*}\centering
\scriptsize
\setlength{\tabcolsep}{5.5pt}
\rowcolors{6}{gray!10}{white}
\begin{tabular}{lrrr|rrr|rrr|rrr}
\toprule
FI Strat. & \multicolumn{3}{c}{\geth} & \multicolumn{3}{c}{\besu} & \multicolumn{3}{c}{\erigon} & \multicolumn{3}{c}{\nethermind} \\
\cmidrule(lr){2-4} \cmidrule(lr){5-7} \cmidrule(lr){8-10} \cmidrule(lr){11-13}
& Available & Degraded & Unavailable & Available & Degraded & Unavailable & Available & Degraded & Unavailable & Available & Degraded & Unavailable \\
\midrule
FI 1 & \score 1.0000 & 0.0000 & 0.0000 & 0.9974 & 0.0025 & 0.0001 & 0.9676 & 0.0324 & 0.0000 & 0.9993 & 0.0000 & 0.0007 \\
FI 2 & \score 1.0000 & 0.0000 & 0.0000 & 0.9999 & 0.0000 & 0.0001 & 0.9703 & 0.0297 & 0.0000 & 0.9993 & 0.0000 & 0.0007 \\
FI 3 & \score 1.0000 & 0.0000 & 0.0000 & 0.9999 & 0.0000 & 0.0001 & 0.9023 & 0.0977 & 0.0000 & 0.9950 & 0.0041 & 0.0010 \\
FI 4 & \score 1.0000 & 0.0000 & 0.0000 & 0.9999 & 0.0000 & 0.0001 & 0.9660 & 0.0340 & 0.0000 & 0.9987 & 0.0000 & 0.0013 \\
FI 5 & \score 1.0000 & 0.0000 & 0.0000 & 0.9937 & 0.0062 & 0.0001 & 0.9780 & 0.0220 & 0.0000 & 0.9990 & 0.0000 & 0.0010 \\
FI 6 & \score 1.0000 & 0.0000 & 0.0000 & 0.9912 & 0.0087 & 0.0001 & 0.9476 & 0.0524 & 0.0000 & 0.9991 & 0.0000 & 0.0009 \\
FI 7 & \score 1.0000 & 0.0000 & 0.0000 & 0.9610 & 0.0269 & 0.0121 & 0.9605 & 0.0395 & 0.0000 & 0.0487 & 0.9392 & 0.0121 \\
FI 8 & \score 1.0000 & 0.0000 & 0.0000 & 0.9882 & 0.0000 & 0.0118 & 0.9625 & 0.0375 & 0.0000 & 0.0003 & 0.9875 & 0.0122 \\
FI 9 & \score 1.0000 & 0.0000 & 0.0000 & 0.9842 & 0.0038 & 0.0120 & 0.0003 & 0.9997 & 0.0000 & 0.0417 & 0.9462 & 0.0121 \\
FI 10 & \score 1.0000 & 0.0000 & 0.0000 & 0.9843 & 0.0037 & 0.0120 & 0.0708 & 0.9292 & 0.0000 & 0.0361 & 0.9515 & 0.0124 \\
FI 11 & \score 1.0000 & 0.0000 & 0.0000 & 0.9253 & 0.0573 & 0.0175 & 0.0580 & 0.9420 & 0.0000 & 0.0422 & 0.9453 & 0.0125 \\
FI 12 & \score 1.0000 & 0.0000 & 0.0000 & 0.9802 & 0.0024 & 0.0174 & 0.0002 & 0.9998 & 0.0000 & 0.0361 & 0.9519 & 0.0120 \\
FI 13 & \score 1.0000 & 0.0000 & 0.0000 & 0.9751 & 0.0076 & 0.0173 & 0.1062 & 0.8938 & 0.0000 & 0.0542 & 0.9339 & 0.0119 \\
FI 14 & \score 1.0000 & 0.0000 & 0.0000 & 0.9833 & 0.0000 & 0.0167 & 0.0002 & 0.9998 & 0.0000 & 0.1236 & 0.8638 & 0.0125 \\
FI 15 & \score 0.9996 & 0.0000 & 0.0004 & 0.9546 & 0.0277 & 0.0176 & 0.2180 & 0.7797 & 0.0023 & 0.0385 & 0.9488 & 0.0127 \\
FI 16 & \score 0.9997 & 0.0000 & 0.0003 & 0.9395 & 0.0427 & 0.0178 & 0.0002 & 0.9994 & 0.0004 & 0.0540 & 0.9338 & 0.0123 \\
FI 17 & 0.3627 & 0.5442 & 0.0931 & \score 0.6556 & 0.2482 & 0.0962 & 0.6183 & 0.2911 & 0.0906 & 0.4507 & 0.2602 & 0.2891 \\
FI 18 & 0.1374 & 0.7643 & 0.0982 & \score 0.6224 & 0.2851 & 0.0925 & 0.5119 & 0.3698 & 0.1183 & 0.1723 & 0.0883 & 0.7393 \\
FI 19 & 0.2628 & 0.6428 & 0.0944 & \score 0.6646 & 0.2308 & 0.1046 & 0.5157 & 0.3778 & 0.1066 & 0.0667 & 0.0210 & 0.9124 \\
FI 20 & 0.2102 & 0.6872 & 0.1026 & \score 0.6254 & 0.2887 & 0.0859 & 0.5430 & 0.3371 & 0.1198 & 0.4207 & 0.3316 & 0.2477 \\
\midrule
Avg. & 0.8486 & 0.1319 & 0.0195 & \score 0.9113 & 0.0621 & 0.0266 & 0.5149 & 0.4632 & 0.0219 & 0.3788 & 0.5053 & 0.1159 \\
\bottomrule
\end{tabular}
\caption{single-version deployments' availability rate for workload B, with varying fault injection (FI) strategies.
The node version with the highest availability rate for the FI row is marked with (\score).}
\label{tab:rpc-availability-single-getblock}
\end{table*}


\begin{mdframed}[style=mpdframe, nobreak=true, frametitle=Answer to RQ1]
Unstable execution environments disrupt the behavior of blockchain nodes in the form of connection issues or broken responses: resets, timeouts, invalid checksums, malformed HTTP or JSON data, etc.
These effects depend on the node version, some effects are observed in all versions, while others are version-specific.
This validates the core assumption of N-Version blockchain nodes: not all sub-nodes will fail in the same way at the same time in an unstable environment.
\end{mdframed}

\subsection{\rqtwo}
\label{sec:results-2}

\autoref{tab:rpc-availability-single-getblock} shows the availability rate of all tested node versions while executing Workload~B and under all fault injection strategies (FI).
Each row corresponds to one FI and the resulting availability rates of each node, the columns correspond to the availability states described in \autoref{sec:availability}.
Regarding full availability, it can be observed that fault injection affects the blockchain nodes in different degrees.
There is a pattern where the nodes can handle increasing aggressiveness of the FIs up to a certain point.
This pattern is consistent with our way of constructing fault injection strategies by increasing aggressiveness.
The first FIs contain the most common system-call errors, and our observations confirm that they are also better handled.
The last FIs use rare and potent system-call errors and put more pressure on nodes' availability.
Nonetheless, the first noticeable degradation varies between nodes:
\geth is able to keep  high availability under the first 16 FIs, and \nethermind and \besu under the first 6.
\erigon presents a different pattern where full availability is slightly disrupted even by the first FI.
Regarding degraded availability, we identify that its main source is the disruption of the deployments' live synchronization under the most aggressive FIs.
This results in responses that do not fulfill the freshness property.
Regarding full unavailability, we do not identify any global pattern other than correlation to the aggressiveness of the FIs.
Additionally, for all node versions, FI~17 causes a sharp increase in unavailability. 

The results also capture a diversity of effects on the nodes given the same FI.
For example, all clients show contrasting behavior under FI~15: while \geth is almost always available, \besu and \nethermind are mostly degraded, and \erigon shows intermittently available behavior.
Overall, the data shows that \besu has the highest availability rate in average, followed by \geth, \erigon, and \nethermind respectively.
Nonetheless, \geth has the lowest average for unavailability, managing to reply in average to $98.1$\% (only $1.9$\% unavailable) of the requests with either available or degraded responses.
Furthermore, \nethermind significantly underperforms the rest of the nodes when full and degraded availability are combined, under the four most aggressive FIs.

The difference in how the fault injection strategies affect the nodes' availability can be explained by the distinct error handling paradigms of the underlying programming stacks.

\begin{mdframed}[style=mpdframe, nobreak=true, frametitle=Answer to RQ2]
The availability of blockchain nodes deteriorates noticeably under unstable execution environments.
The measure at which this happens varies depending on the node version and fault injection strategy, in average:
\geth's availability drops to $0.8486$;
\besu's availability drops to $0.9113$;
\erigon's availability drops to $0.5149$;
and \nethermind's availability drops to $0.3788$.
All node versions remain available in certain conditions where the others become unavailable.
In other words, none of the tested fault injection strategies make all nodes unavailable simultaneously.
Now, we have strong evidence that the available diverse blockchain node versions are suitable for our novel N-Version design.
\end{mdframed}

\subsection{\rqthree}

\autoref{tab:rpc-availability-N-Version-getblock} shows the availability measurement of our N-Version blockchain node prototype \neth under unstable environment while executing workload B.
It shows the best performing combinations in average given $N$ values of 2, 3, and 4.
With $N=2$, and executing \geth and \erigon, \neth is able to maintain $94.2$\% full availability, and $99.9978$\% combined full and degraded availability.
When $N=3$, with \geth, \besu, and \erigon, \neth is able to maintain $97.3$\% full availability, and $99.98$\% combined full and degraded availability.
With $N=4$, and executing \geth, \besu, \erigon, and \nethermind, \neth is able to maintain $98.5$\% full availability, and $99.9999$\% combined full and degraded availability.
Similar to the results presented in \autoref{sec:results-2}, \neth is able to perform normally under the first 16 fault injection strategies.
For FI~19, the only strategy with imperfect mitigation, only $0.02$\% of the requests result in an unavailable response.

\begin{table*}\centering
\scriptsize
\rowcolors{4}{gray!10}{white}
\begin{tabular}{lrrrrrrrrrr}
    \toprule
FI Strat. & \multicolumn{3}{c}{\neth-2 (\textsc{Ge.} + \textsc{Er.})} & \multicolumn{3}{c}{\neth-3 (\textsc{Ge.} + \textsc{Be.} + \textsc{Er.})} & \multicolumn{3}{c}{\neth-4} \\
\cmidrule(lr){2-4} \cmidrule(lr){5-7} \cmidrule(lr){8-10}
&  \multicolumn{1}{c}{Available} &  \multicolumn{1}{c}{Degraded} &  \multicolumn{1}{c}{Unavailable} &  \multicolumn{1}{c}{Available} &  \multicolumn{1}{c}{Degraded} &  \multicolumn{1}{c}{Unavailable} &  \multicolumn{1}{c}{Available} &  \multicolumn{1}{c}{Degraded} &  \multicolumn{1}{c}{Unavailable} \\
\midrule
FI 1 & 1.0000 & 0.0000 & 0.0000 & 1.0000 & 0.0000 & 0.0000 & 1.0000 & 0.0000 & 0.0000 \\
FI 2 & 1.0000 & 0.0000 & 0.0000 & 1.0000 & 0.0000 & 0.0000 & 1.0000 & 0.0000 & 0.0000 \\
FI 3 & 1.0000 & 0.0000 & 0.0000 & 1.0000 & 0.0000 & 0.0000 & 1.0000 & 0.0000 & 0.0000 \\
FI 4 & 1.0000 & 0.0000 & 0.0000 & 1.0000 & 0.0000 & 0.0000 & 1.0000 & 0.0000 & 0.0000 \\
FI 5 & 1.0000 & 0.0000 & 0.0000 & 1.0000 & 0.0000 & 0.0000 & 1.0000 & 0.0000 & 0.0000 \\
FI 6 & 1.0000 & 0.0000 & 0.0000 & 1.0000 & 0.0000 & 0.0000 & 1.0000 & 0.0000 & 0.0000 \\
FI 7 & 1.0000 & 0.0000 & 0.0000 & 1.0000 & 0.0000 & 0.0000 & 1.0000 & 0.0000 & 0.0000 \\
FI 8 & 1.0000 & 0.0000 & 0.0000 & 1.0000 & 0.0000 & 0.0000 & 1.0000 & 0.0000 & 0.0000 \\
FI 9 & 1.0000 & 0.0000 & 0.0000 & 1.0000 & 0.0000 & 0.0000 & 1.0000 & 0.0000 & 0.0000 \\
FI 10 & 1.0000 & 0.0000 & 0.0000 & 1.0000 & 0.0000 & 0.0000 & 1.0000 & 0.0000 & 0.0000 \\
FI 11 & 1.0000 & 0.0000 & 0.0000 & 1.0000 & 0.0000 & 0.0000 & 1.0000 & 0.0000 & 0.0000 \\
FI 12 & 1.0000 & 0.0000 & 0.0000 & 1.0000 & 0.0000 & 0.0000 & 1.0000 & 0.0000 & 0.0000 \\
FI 13 & 1.0000 & 0.0000 & 0.0000 & 1.0000 & 0.0000 & 0.0000 & 1.0000 & 0.0000 & 0.0000 \\
FI 14 & 1.0000 & 0.0000 & 0.0000 & 1.0000 & 0.0000 & 0.0000 & 1.0000 & 0.0000 & 0.0000 \\
FI 15 & 0.9996 & 0.0004 & 0.0000 & \triup1.0000 & 0.0000 & 0.0000 & 0.9999 & 0.0001 & 0.0000 \\
FI 16 & 0.9997 & 0.0003 & 0.0000 & \triup1.0000 & 0.0000 & 0.0000 & \triup1.0000 & 0.0000 & 0.0000 \\
FI 17 & 0.7377 & 0.2547 & 0.0076 & 0.8424 & 0.1568 & 0.0009 & \triup0.9511 & 0.0489 & 0.0000 \\
FI 18 & 0.6042 & 0.3846 & 0.0112 & 0.8587 & 0.1392 & 0.0021 & \triup0.9891 & 0.0109 & 0.0000 \\
FI 19 & 0.7626 & 0.2244 & 0.0131 & \triup0.9254 & 0.0745 & 0.0001 & 0.8354 & 0.1645 & 0.0002 \\
FI 20 & 0.7383 & 0.2495 & 0.0122 & 0.8470 & 0.1525 & 0.0005 & \triup0.9210 & 0.0790 & 0.0000 \\
\midrule
Avg. & 0.9421 & 0.0557 & 0.0022 & 0.9737 & 0.0261 & 0.0002 & \triup0.9848 & 0.0152 & 0.0000 \\
\bottomrule
\end{tabular}

\caption{N-Version node availability for workload B, with varying N and fault injection (FI) strategies.
The arrows represent the change of the rates compared to the best single-version node. The table shows only the combinations with the highest availability in average for each N. (\triup) indicates the value where the highest gain in availability was achieved.}
\label{tab:rpc-availability-N-Version-getblock}
\end{table*}

\begin{table*}\centering
\scriptsize
\rowcolors{4}{gray!10}{white}
\begin{tabular}{llrrrrrr}
    \toprule
{} & \multicolumn{1}{c}{\multirow{2}{*}{Combination}} & \multicolumn{3}{c}{Available + Degraded} & \multicolumn{3}{c}{Resource usage} \\
\cmidrule(lr){3-5} \cmidrule(lr){6-8}
{} & {} & \multicolumn{1}{c}{FI 18} & \multicolumn{1}{c}{FI 19} & \multicolumn{1}{c}{FI 20} & \multicolumn{1}{c}{CPU  \%} & \multicolumn{1}{c}{RAM (GBs)} & \multicolumn{1}{c}{Disk (GBs)} \\
\midrule
\cellcolor{white} & \textsc{Ge.} + \textsc{Be.} & 0.9797 & 0.9740 & 0.9998 & 338.90 & 119.07 & 1850 \\
\cellcolor{white} & \textsc{Ge.} + \textsc{Er.} & 0.9888 & 0.9869 & 0.9878 & 370.24 & 355.92 & 2294 \\
\cellcolor{white} & \textsc{Ge.} + \textsc{Ne.} & 0.9636 & 0.9616 & 0.9247 & 314.49 & 113.00 & 2246 \\
\cellcolor{white} & \textsc{Be.} + \textsc{Er.} & 0.9994 & 0.9736 & 0.9691 & 269.24 & 267.51 & 2350 \\
\cellcolor{white} & \textsc{Be.} + \textsc{Ne.} & 0.9789 & 0.9892 & 0.9435 & 213.49 & 24.59 & 2302 \\
\cellcolor{white}\multirow{-6}{*}{\shortstack{\neth \\ N=2}} & \textsc{Er.} + \textsc{Ne.} & 0.9577 & 0.9351 & 0.9936 & 244.84 & 261.44 & 2746 \\
\midrule
\cellcolor{white} & \textsc{Ge.} + \textsc{Be.} + \textsc{Er.} & 0.9979 & 0.9999 & 0.9995 & 489.19 & 371.25 & 3247 \\
\cellcolor{white} & \textsc{Ge.} + \textsc{Be.} + \textsc{Ne.} & 0.9997 & 0.9914 & 0.9994 & 433.44 & 128.33 & 3199 \\
\cellcolor{white} & \textsc{Ge.} + \textsc{Er.} + \textsc{Ne.} & 1.0000 & 0.9922 & 1.0000 & 464.79 & 365.18 & 3643 \\
\cellcolor{white}\multirow{-4}{*}{\shortstack{\neth \\ N=3}} & \textsc{Be.} + \textsc{Er.} + \textsc{Ne.} & 0.9995 & 0.9724 & 0.9999 & 363.79 & 276.77 & 3699 \\
\midrule
\cellcolor{white}\shortstack{\neth \\ N=4} & \textsc{Ge.} + \textsc{Be.} + \textsc{Er.} + \textsc{Ne.} & 1.0000 & 0.9998 & 1.0000 & 583.74 & 380.51 & 4596 \\
\bottomrule
\end{tabular}

\caption{N-ETH configurations and their respective Available + Degraded scores, and resource usage measurement. (\textsc{Ge.}) \geth, (\textsc{Be.}) \besu, (\textsc{Er.}) \erigon, (\textsc{Ne.}) \nethermind. This table only presents the scores achieved under the 3 most aggressive FIs. Resource usage is measured under normal execution.}
\label{tab:rpc-availability-vs-resources}
\end{table*}

By comparing \neth's rates with single-version's rates, we see that these are equal or improved under all fault injection strategies.
This is, the 'available', 'degraded', and 'unavailable' rates are either better than or equal to the best single-version node for all fault injection strategies.
More specifically, in the most aggressive 4 FIs, both the 'available' and 'unavailable' rates are strictly better.
For instance, in the single-node deployments, \besu is the best at handling injection FI~20, with scores: $0.6254$ available, $0.2887$ degraded, and unavailable $0.0859$.
In contrast, \neth with $N=4$ handles the same fault injection strategy with significantly better scores: $0.9210$ available, $0.0790$ degraded, and unavailable $0$.
This represents a difference in scores of: $+0.2956$ available, $-0.2097$ degraded, and $-0.0859$ unavailable. In summary, this is close to $50$\% more available.

The 'unavailable' rate of \neth drops to zero or close to zero with the increasing $N$, for all fault injection strategies.
This indicates that the N-Version blockchain node sustains at least a degraded service for most of the tested unstable execution environments.
It is important to note that the most noticeable gain is achieved for the most aggressive FIs (FI~17-20).
For instance, during FI~18, \besu's unavailable score is $0.0925$, while \neth with $N$ values of $2$, $3$, and $4$ drop to $0.0112$, $0.0021$, and $0$ respectively.

Under fault injection strategies FI~15 and 19, \neth with $N=4$ is slightly worse than \neth with $N=3$.
This can be attributed to two main factors.
First, the fault injection tool is non-deterministic, \ie it injects system call errors based on the probabilities defined by the fault injection strategies.
Second, \neth's sub-nodes are continuously being updated with the latest blocks, \ie each experiment is carried out using Ethereum's production state.
This is a deliberate experiment design decision, as it allows us to measure changes in availability in the current real-world environment.
Given these two points, the injected faults across experiments are never performed exactly with the same underlying state.
We mitigate this difference across experiments by performing \numprint{360000} requests during each of them.

\autoref{tab:rpc-availability-vs-resources} shows the availability scores of all possible combinations of \neth given $N=2$, $3$, and $4$.
We  observe that the increase of availability is correlated to $N$, and  increasing $N$ also increases resource usage, as expected.
These results show that there are differences in trade-offs in the combination space.
For instance, the combination of \geth and \besu is significantly more available than the combination of \geth and \nethermind.
In these cases, during the experiment using FI~20, full and degraded combined availability rates are $99.98$\% and $92.47$\% respectively.
\autoref{tab:rpc-availability-vs-resources} also shows the variability of resource usage from the different combinations.
For example, with $N=2$ the combinations vary by an order of magnitude in terms of RAM. 
This information is useful for users who want to select sub-node combinations with $N<4$ for sake of resource constraints or other reasons.

Overall, these experimental results demonstrate that N-Version blockchain nodes provide higher availability than single-version nodes under the same unstable execution environments.

\begin{mdframed}[style=mpdframe, nobreak=true, frametitle=Answer to RQ3]
N-Version blockchain nodes have better availability compared to single-version blockchain nodes.
Our data shows that the gains are significant, especially for aggressive fault injection scenarios.
As compared to single-version blockchain nodes, we observe in average an increase in availability from  $84.7$\%, up to $98.5$\%.
Notably, the $N=4$ deployment reduces full unavailability to a negligible amount.
These results validate the overall usefulness of our novel use of N-Version design in the context of blockchains.
Our results are of utmost importance for practitioners who either provide (\eg Infura) or rely on blockchain nodes (\eg exchanges, banks, and art platforms).
\end{mdframed}

\section{Discussion}

\subsection{Overhead and trade-offs}
The main trade-off of N-Version design is the enhancement of a desired property versus increased resource usage.
In the case of \neth, availability under instability is enhanced significantly at the expense of an increase of computing resources, dictated by the number of versions $N$, as shown by our results.
Yet, access to computing resources is not a major issue for large service providers, since they are already used to run a sizable amount of blockchain node instances~\cite{li2021strong}.
Those providers would greatly benefit from higher availability and automatic resilience to hardware or software-related faults.

\subsection{Threats to validity}
\textit{Threats to internal validity:}
We identify two sources of noise that can have an effect on the produced data and corresponding findings.
First, we use non-deterministic fault injection strategies, which can trigger system-calls at any time of the experiment, and during different stages of execution.
Second, we use fully-synchronized blockchain nodes, which implies that every experiment is performed over a blockchain node that uses a different underlying state. 
We mitigate both sources of noise by using a large number of requests in each of the experiments.

\textit{Threats to external validity:}
We identify two stages where the obtained data produces generalizable results:
First, we generalize the availability metrics obtained from one RPC method (Workload B), to the whole API.
We argue that this is realistic, since the effects of unstable execution environments are uniform across the API.
External validity would be improved by considering write methods of the API, this is considered as future work.
Second, we choose to realize our prototype implementation in Ethereum, as it is a popular, actively supported, and mature blockchain.
We argue that our results are generalizable to other blockchain technologies, as they follow generally the same design principles, yet this has to be verified empirically.

\section{Related Work}

\subsection{Blockchain Dependability}
Kolb~\etal~\cite{Kolb2020} survey open challenges regarding blockchain technology, including the need to enhance non-functional properties such as scalability and availability.
Weber~\etal~\cite{Weber2017} present a thorough analysis of the availability of major blockchains, and conclude that while read availability is typically high, write availability is low due to uncommitted transactions.
While these studies address blockchain availability, they do not account for the effects of unstable execution environments.

In the field of Chaos Engineering, Ma et al.~\cite{phoenix} present Phoenix, a system to detect resilience issues in blockchains using context-sensitive fault injection. In contrast to our work, the focus of Phoenix is to discover the causes of unrecoverable states in blockchain nodes.

The state of client diversity in Ethereum is described by Ranjan~\cite{Pooja2020}, and is continuously tracked by the Ethereum community~\cite{ClientDiversity}.
In the area of dependability through diversity, Garcia~\etal~\cite{Garcia2019} present Lazarus, a tool for automatic management of diversity in Byzantine fault-tolerant systems.
Breidenbach~\etal~\cite{Breidenbach2018} present the Hydra framework, whose goal is to enhance security of smart contracts using N-Version programming.
These works study dependability in closely related application domains, however their specific focus is different.
Lazarus~\cite{Garcia2019} focuses on systemic-level dependability of BFT systems; and Hydra in dependability of smart contracts.
In contrast, this work is focused on the external availability of nodes of blockchain systems.

Regarding security of blockchains, Chen~\etal~\cite{Chen2020} outline the security of the Ethereum ecosystem, by detailing vulnerabilities, attacks, and defenses.
Groce~\etal~\cite{Groce2020:FlawsInContracts} invited 23 professional stakeholders to audit Ethereum smart contracts using both tools and manual analysis, with 246 individual defects identified, categorized based on their severity and difficulty.
Groce~\etal~\cite{Groce2022} investigate weaknesses in the Bitcoin Core's fuzzing project.
While these works address security, a fundamental attribute of blockchain dependability, availability is not their main consideration.

In the work of Li~\etal~\cite{deter, li2021strong}, the effects of certain DoS attacks are measured at the node level.
Likewise, Yang~\etal~\cite{Yang2021} and Kim et al.~\cite{etherdiffer} perform differential testing on Ethereum nodes.
Their effort led to the discovery of several bugs in the target nodes, which greatly contributed to strengthening the Ethereum blockchain.
However, these works' focus is different from ours, namely security, consensus reliability, and response consistency.

\subsection{Software Diversity}
Multi-version approaches such as N-Version programming and N-variant systems have been extensively researched and proven to enhance security and fault tolerance~\cite{Baudry2015}.
Seminal work from Avizienis~\etal~\cite{Avizienis1985} introduced N-Version programming, which highlights the opportunities of diverse computation for making fault-tolerant systems.
Theoretical analyses and models of N-Version software~\cite{eckhardtandlee, littlewood} agree that independence of behavior is crucial for achieving fault-tolerance goals.
These works also agree that in practice this independence cannot be guaranteed, even if the versions are developed by distinct teams or using distinct methodologies~\cite{knightandleveson}.
Therefore, the applicability of N-Version software has been empirically studied over an expansive range of domains.
Within this range we highlight the works of Xue~\etal~ on web browsers \cite{DBLP:conf/ndss/XueDK12}, Chauvel~\etal~ on cloud architecture \cite{Chauvel15}, Huang~\etal~ on web services~\cite{huang2011introducing}, Oberheide~\etal~on anti-viruses~\cite{cloudAV}, and Harrand~\etal~on Java libraries~\cite{harrand2021automatic}, because of their use of natural diversity, as opposed to planned diversity in Avizienis' vision.
Similar to \neth, the mentioned works leverage domain knowledge to achieve fault-tolerance, availability, and security.
However, to the best of our knowledge, this is the first work to propose natural diversity and N-Version design as a means of hardening blockchain infrastructure.

N-Version programming traditionally relies on majority voting to select a response to a request; however, several alternatives have been presented.
For example, Vouk~\etal~\cite{Vouk1993} proposes consensus voting, where a response in an N-Version system is selected only after $M < N/2$ versions agree to accept a response.
Going beyond voting, Gao~\etal~\cite{DBLP:journals/tdsc/GaoRS09} describe the use of Hidden Markov Models to compute the low-level behavioral distance of versions for anomaly detection.
In our work, we propose selecting a final response after only one timely, compliant, and fresh response is produced from any subnode.
This approach allows \neth to provide high availability.

A similar, but more security-focused concept is N-variant systems~\cite{Cox2006, Franz2018}.
In contrast to N-Version programming, “variants” are automatically generated.
Koning and colleagues propose \textsc{MvArmor}, an N-variant execution engine that exploits hardware virtualization to detect divergent behavior among program variants~\cite{koning2016secure}, where behavior divergence is observed at the level of system-calls.
Voulimeneas~\etal~\cite{DBLP:conf/dimva/VoulimeneasSPNL20} show how N-variant execution can be based on the diversity of Instruction Set Architectures by running programs natively in different machines.
Berger~\etal~\cite{berger2006diehard} propose a runtime system to handle errors through diversified memory layouts.
Polinsky~\etal~\cite{polinsky2020nm} propose an extension to N-M-variant systems, where M represents a number of replicas for each variant and guarantees a constant N throughout a period where a variant's instance might be unavailable.
N-variant systems increase diversity at low levels in single software stacks.
As such, they do not mitigate flaws originating from application design, dependencies, or programming languages, which is \neth's explicit goal.
Nonetheless, the mentioned N-variant approaches and \neth are not mutually exclusive.
These can potentially be used in combination, providing an even greater spectrum for resilience by diversity.

The proxy pattern and N-Version design are often studied in combination.
For instance, Espinoza~\etal~\cite{DBLP:conf/dsn/EspinozaWFT22} describe a design where N-Versioned microservices are placed in between proxies, allowing the system to compare both upstream and downstream request-response pairs.
Simillarly, Durieux~\etal~\cite{durieux2020fully} leverage protocol diversity through an HTTP proxy to introduce self-healing for HTML and JavaScript code.
These works show that diversity together with proxies can be used to augment targeted aspects of software, \eg mitigating security concerns or providing automatic code repair.
In the intersection of blockchains and N-Version design, proxies are addressed in smart contract resilience.
For instance, The Hydra framework~\cite{Breidenbach2018}, describes the entry point of their N-Version smart contracts as a "generic proxy" which delegates incoming transactions to each version.
Similarly, Péter~\etal~\cite{peter2023n} propose a proxy between N-Versioned smart contracts and the underlying storage of the Hyperledger Fabric blockchain.
These works show that the proxy pattern can be applied at distinct layers of blockchain systems.
However, none suits \neth's problem statement: high availability in unstable environments. 
These previous works are different from \neth's original solution of RPC routing, response selection and adaptive proxying tuned for blockchain nodes.

\section{Conclusion and Future Work}

In this paper, we identify the potential of taking advantage of existing diversity of blockchain node implementations.
We devise an architecture that aims to improve the availability of blockchain clients under suboptimal, unstable execution.
We implement a prototype based on this design: \neth, and evaluate its availability against regular blockchain nodes.
To simulate unstable execution environments, we use a system-call error injection tool.

Our findings show that:
(1) External applications which consume blockchain nodes' APIs perceive erratic behavior when the target node is under unstable execution environments;
(2) The availability of blockchain nodes is affected by the tested unstable execution environments.
The severity of the effects in availability scales with the aggressiveness of the used fault injection strategies; and
(3) The N-Version blockchain node prototype is able to stay in available or degraded state under most of the tested unstable execution environments.
Additionally, \neth presents a drastic reduction in unavailability when compared to common blockchain nodes, which present much larger unavailability windows under the same unstable execution environments.
Ultimately, this is the benefit of relying on strong versions that have different weaknesses, as \neth mitigates the failures surfacing on specific node version and fault scenario combinations.

In the presented architecture and prototype, we focus on blockchain node implementation diversity.
However, we can identify two other dimensions where diversity is relevant and applicable.
First, \textit{operating system diversity}, where blockchain nodes are executed on top of diverse operating systems.
This approach has the potential to enhance OS-related fault tolerance and security.
Second, \textit{single node diversity}, where different versions of the same node are used to detect regressions or errors introduced in newer versions.

In this paper, we focus on improving on availability, which is business critical to external clients and applications.
Yet, we envision that N-Version blockchain nodes can enhance dependability attributes other than availability, such as reliability and security.
Furthermore, it can be used to enhance performance metrics perceived by external clients such as latency and throughput, given that different blockchain nodes are based on competing design principles.

\hfill

\section{Acknowledgements}
This work was supported by the CHAINS project funded Swedish Foundation for Strategic Research (SSF), as well as by the Wallenberg Autonomous Systems and Software Program (WASP) funded by the Knut and Alice Wallenberg Foundation.

\bibliographystyle{unsrt}
\bibliography{references}

\end{document}